\documentclass[aps,prd,nofootinbib]{revtex4}
\usepackage[colorlinks=true, pdfstartview=FitV, linkcolor=blue, citecolor=red, urlcolor=magenta]{hyperref}
\usepackage{graphicx}
\usepackage{latexsym}
\usepackage{amsmath}
\usepackage{amsfonts}
\usepackage{amssymb}
\usepackage{verbatim}

\newcommand{\be}{\begin{equation}}
\newcommand{\ee}{\end{equation}}
\newcommand{\bea}{\begin{eqnarray}}
\newcommand{\eea}{\end{eqnarray}}

\newcommand{\pa}{\partial}

\newcommand{\bb}{\bibitem}

\def\bb{\bibitem}

\def\g{\gamma}

\def\bb{\bibitem}
\newcommand{\ben}{\begin{eqnarray}}
\newcommand{\een}{\end{eqnarray}}

\begin{document}

\title{Lorentz invariance violation and simultaneous emission of electromagnetic and gravitational waves}
\author{$^{1,3}$E. Passos}
\email{passos@df.ufcg.edu.br}
\author{$^{1}$M. A. Anacleto}
\email{anacleto@df.ufcg.edu.br}
\author{ $^{1,2}$F. A. Brito}
\email{fabrito@df.ufcg.edu.br}
\author{$^{4}$O.  Holanda}
\email{ozorio.neto@uerj.br}
\author{$^{1}$G. B. Souza}
\email{gilvancx.fis@hotmail.com}
\author{$^{3}$C. A. D. Zarro}
\email{carlos.zarro@if.ufrj.br}

\affiliation{$^{1}$Departamento de F\'{\i}sica, Universidade Federal de Campina Grande,\\
Caixa Postal 10071, 58429-900, Campina Grande, Para\'{\i}ba, Brazil.}
\affiliation{$^{2}$Departamento de F\' isica, Universidade Federal da Para\' iba,\\  Caixa Postal 5008, Jo\~ ao Pessoa, Para\' iba, Brazil.}
\affiliation{$^{3}$Instituto de F\' isica, Universidade Federal do Rio de Janeiro,\\  Caixa Postal 21945, Rio de Janeiro, 
Rio de Janeiro, Brazil.}
\affiliation{$^{4}$Departamento de F\'{i}sica Te\' orica, Instituto de F\' isica, Universidade do Estado do Rio de Janeiro,
Rua S\~ ao Francisco Xavier 524,  20550-013,  Maracan\~ a,  Rio de Janeiro, Brazil.}


\begin{abstract}
In this work, we compute some phenomenological bounds for the electromagnetic and massive gravitational high-derivative extensions supposing that it is possible to have an astrophysical process that generates simultaneously gravitational and electromagnetic waves. We present Lorentz invariance violating (LIV) higher-order derivative models, following the Myers-Pospelov approach, to electrodynamics and massive gravitational waves. We compute the corrected  equation of motion of these models, their dispersion relations and the velocities. The LIV parameters for the gravitational and electromagnetic sectors, $\xi_{g}$ and $\xi_{\gamma}$, respectively, were also obtained for three different approaches: luminal photons, time delay of flight  and the difference of graviton and photon velocities. These LIV parameters depend on the mass scales where the LIV-terms become relevant, $M$ for the electromagnetic sector and $M_{1}$ for the gravitational one. We obtain, using the values for $M$ and $M_{1}$ found in the literature, that $\xi_{g}\sim10^{-2}$, which is expected to be phenomenologically relevant and $\xi_{\gamma}\sim10^{3}$, which cannot be suitable for an effective LIV theory. However, we show that $\xi_{\gamma}$ can be interesting in a phenomenological point of view if $M\gg M_{1}$. Finally the relation between the variation of the velocities of the photon and the graviton in relation to the speed of light was calculated and resulted in $\Delta v_{g}/\Delta v_{\g} \lesssim 1.82 \times 10^{-3}$.

\end{abstract}
\pacs{XX.XX, YY.YY} \maketitle


\section{Introduction}
The Lorentz invariance violating (LIV) theories have been extensively studied at high energy systems. The main focus is to develop an effective probe to test the phenomenological limits of Lorentz invariance as a direct consequence of the fuzzy nature of spacetime provided by quantum gravity theories on the Planck scale. 
In this context, the possible effects related to LIV are given by energy and helicity dependent photon propagation velocities. The LIV parameters bounds on energy can be inferred by measuring the arrival times of photons with different energies emmited almost simultanously from distant objects \cite{Camelia01}. In order to measure such bounds, it is necessary an ultra high energy phenomena such as a gamma-ray burst (GRB) \cite{Laurent:2011he, Ahmadi:2012we, Krawczynski:2013uga,Abdo} or a flare of an active galactic nucleus \cite{Aharonian,Albert:2007qk}. The LIV parameters can also be constrained by measuring how the polarization direction of an x-ray beam of cosmological origin changes as function of energy \cite{Gambini01}. Such observations have been used as astrophysical laboratories to verify the possible occurrences of LIV in nature \cite{Gleiser, Camal01, Matt, Macc}.

Among several approaches to investigate LIV effective theories, here we shall follow the one proposed by Myers and Pospelov \cite{MP}. In order to break the Lorentz symmetry they introduced mass operators of dimension-five along with a nondynamical four-vector $n_{\mu}$
interacting with scalar, fermion and photon fields. If we restrict our attention only to photon sector, there is a single contribution of dimension-five operators which gives a correction of order $\xi_{\gamma} p^{3}/M_{\rm Pl}$. The extension to dimension-$n$ operators should satisfy all the Myers-Pospelov approach criteria: $(\rm i)$ quadratic in the fields,
$(\rm ii)$ one more derivative than the usual terms, $(\rm iii)$ gauge invariant, $(\rm iv)$ Lorentz invariant, except for the presence of an external four-vector $n_{\mu}$, $(\rm v)$ not reducible to lower dimensional operators at the equations of motion, and $(\rm vi)$ not reducible to a total derivative. In this set-up, one finds that for dimension-$n$ operators, the correction is given by  $\xi_{\gamma} p^{n}/(M_{\rm Pl})^{n-2}$.


The detection of gravitational waves, reported by the  Laser Interferometer Gravitational-Wave Observatory (LIGO) and Virgo Collaborations \cite{GW01,GW02}, opens a new window in observational cosmology  and astrophysics. Particularly in astrophysics, the hitherto detected gravitational waves come from the merger of two black-holes. However, it is expected to also measure gravitational waves emitted during the merger of other compact astrophysics objects such as neutron stars or a black-hole and a neutron star. The merger of such two compact objects is supposed to be a very complex phenomena, probably involving electromagnetic waves or neutrino emission. Hence to obtain new insights of the merger process, one can observe simultaneously the emission of gravitational waves and electromagnetic waves or neutrinos. It was reported, in the event GW150914 the observation of a gravitational wave \cite{GW01} and a short gamma-ray burst (also detected in the same event GW150914 \cite{Fermi01}) by the Fermi Gamma-Ray Space Telescope. These observations have been used to obtain constraints on LIV parameters \cite{Blas, Ellins, Gia,  Vincenzo} (see also \cite{Will}). This issue started an intense debate in the literature as some authors describe that this electromagnetic counterpart is not possible \cite{BH01} and others showed its plausibility \cite{BH02,Morsony:2016upv}. The null results of simultaneous GRB emission and the other detected gravitational wave event \cite{Smartt:2016oeu,Yoshida:2016ddu}, GW151226 \cite{GW02}, apparently favours that  this simultaneous emission was unlikely, however it is not possible to conclude anything, as there are only two detected gravitational waves events and the physical processes involved in electromagnetic waves and neutrino emissions are not entirely understood.


A transient GRB signal above $50$ keV after 0{.}40 s after the detection of GW150914 was reported in Ref. \cite{Fermi01}. This observational fact and a bound for the graviton mass, $m_{g}\leq10^{-22}$  eV, was used in Refs. \cite{Blas, Ellins, Gia,  Vincenzo} to obtain bounds constraining the difference between light and graviton speed and the energy scale where the LIV effects appear. Our goal is to extend these references by introducing  high-order derivative operators which explicitly break Lorentz symmetry. The main purpose of this work is to consider the electromagnetic and gravitational dispersion relations produced by  presence of the higher derivative operator in the effective actions. We aim to find new phenomenological constraints on LIV by using simultaneous measures of the gamma-ray busts (GRB) and gravitational wave (GW) produced by the same source, i.e, assuming that such signals emerge from black holes merger.

The outline of this paper is as follows. In Sec. \ref{sec02}, we have demonstrated that both the electromagnetic higher derivative term appear as terms of a power series associate to CPT-odd effective actions.  We use a dimension-five operator as modification to Maxwell Lagrangian. The associated dispersion relations are obtained. In Sec. \ref{sec03}, we reproduces the electromagnetic calculations to linearized gravity. We also use a dimension-five operator to modify the massive Fierz-Pauli Lagrangian. The associated dispersion relations are obtained for massive and massless cases. In Sec. \ref{sec04}, we discussed some phenomenological constraints. In Sec.\ref{sec05}, we present our conclusions.

\section{The higher derivative LIV extensions: electromagnetic sector}\label{sec02}


\subsection{The Model}
Now, we consider a CPT-odd pure-photon action proposed by  Carrol-Field-Jackiw (CFJ) \cite{CFJ} and through it we get higher derivative contributions of a power series.  The CFJ action is rewritten as 
\bea\label{cfj1}
S_{\gamma}&=& -\frac{M \xi_{\gamma}}{2}\int d^{4}x\,f^{\mu\lambda\nu}F_{\lambda\mu}A_{\nu},\;\; f^{\mu\lambda\nu}\equiv\varepsilon^{\alpha\mu\lambda\nu}n_{\alpha}\nonumber\\&=&
\frac{M \xi_{\gamma}}{2}\int d^{4}x\,A_{\mu}\Pi^{\mu\nu}A_{\nu}
\eea
where $\xi_{\gamma}$ is a dimensionless parameter, $M$ is the mass scale where LIV effects emerges and $\Pi^{\mu\nu}= f^{\mu\lambda\nu}\pa_{\lambda}$ is the electromagnetic LIV operator that enjoys the following properties: 
$\pa_{\mu}\Pi^{\mu\nu}= 0,\;\; n_{\mu}\Pi^{\mu\nu}=0,$ $
 \Pi_{\mu\nu}\Pi^{\nu\beta}=
- \big[\delta^{\beta}_{\mu} \big((n\cdot \pa)^{2} - n^{2} \pa^{2}\big) - n^{\beta}\big(\pa_{\mu} (n\cdot \pa) - n_{\mu}\pa^{2}\big)- \pa^{\beta}\big(n_{\mu} (n\cdot \pa) - n^{2}\pa_{\mu}\big)\big]$ and $\Pi_{\mu\nu}\Pi^{\mu\nu}= 2 \big( (n\cdot\pa)^{2} - n^{2}\pa^{2}\big)$. Notice that the above effective action is gauge invariant (up to a surface term) under gauge transformations $\delta A_{\mu}=\pa_{\mu}\Lambda$. 

We now extend the CFJ action by replacing the quantity $\xi_{\gamma} M \Pi^{\mu\nu}$ to a power series such as
\bea\label{cfj2}
\sum_{l=1,3,...}\frac{\xi_{\gamma_{l}}}{(M)^{l - 2}} (\Pi^{\mu\nu})^{l}&=&
\xi_{\gamma_{1}} M \Pi^{\mu\nu} + \frac{\xi_{\gamma_{3}}}{M}\big( \Pi^{\mu\alpha}\Pi_{\alpha\beta}\Pi^{\beta\nu}=\Pi^{\mu\nu}\hat{D}\big)+ ... \nonumber \\
&=& \xi_{\gamma_{1}} M \Pi^{\mu\nu} + \frac{\xi_{\gamma_{3}}}{M}\Pi^{\mu\nu}\hat{D} + ...
\eea
where $\hat{D}= (n\cdot \pa)^{2} - n^{2} \pa^{2}$ is a LIV derivative operator. Inserting the series (\ref{cfj2}) into the action (\ref{cfj1}) we get
\bea\label{cfj3}
S_{\gamma}\to \hat{S}_{\gamma}= -\frac{1}{2}\int d^{4}x\,\Big[M \xi_{\gamma_{1}}f^{\mu\lambda\nu}F_{\lambda\mu}A_{\nu}+ 
\frac{\xi_{\gamma_{3}}}{M} f^{\mu\lambda\nu}\,\hat{D} F_{\lambda\mu}A_{\nu}+...\Big]
\eea
We can rewrite the Eq.~(\ref{cfj3}) in the form
\bea\label{cfj4}
S_{\gamma}\to \hat{S}_{\gamma}= -\frac{1}{2}\int d^{4}x\,\Big[M \xi_{\gamma_{1}}f^{\mu\lambda\nu}F_{\lambda\mu}A_{\nu}+ 
\frac{\xi_{\gamma_{3}}}{M} f^{\mu\lambda\rho} n^{\nu} n^{\rho} (\pa_{\lambda}F_{\mu\nu} )F_{\rho\sigma}-
\frac{\xi_{\gamma_{3}}}{M} f^{\mu\lambda\rho} n^{2} \eta^{\nu\rho} (\pa_{\lambda}F_{\mu\nu} )F_{\rho\sigma}+...\Big].
\eea
Notice that we obtain dimension-five operators as extra terms of power series expansion in Eq.~(\ref{cfj3}) (or Eq.~(\ref{cfj4})) which lead to cubic modifications of the dispersion relations of electromagnetic waves. These contributions obey the main Myers-Pospelov criteria \cite{MP}. {Note also that the above electromagnetic modification can be viewed as the five-dimensional case of  standard model extension photon sector for operators of arbitrary dimension \cite{Kost_Mewes_01}.
}



In this point we consider the second and third terms of Eq.~(\ref{cfj4}) to modify classical electrodynamics (electromagnetic Maxwell-Myers-Pospelov model). In this case we analyze the dispersion relation of electromagnetic waves. 

Let us now derive the dynamics associated with the following Lagrangian
\bea\label{EE01}
{\cal L}_{\gamma} = -\frac{1}{4} F_{\mu\nu}F^{\mu\nu} - \frac{\xi_{\gamma}}{M} f^{\mu\lambda\rho} n^{\nu} n^{\sigma} (\pa_{\lambda}F_{\mu\nu} )F_{\rho\sigma} + \frac{\xi_{\gamma}}{M} f^{\mu\lambda\rho} n^{2} \eta^{\nu\sigma} (\pa_{\lambda}F_{\mu\nu} )F_{\rho\sigma}
\eea
where $\xi_{\gamma_{3}} \equiv \xi_{\gamma} $. Using the axial gauge $n^{\mu} A_{\mu}=0$, the equation of motion reads

\bea\label{EE02}
\big(\pa^{2} \eta^{\mu\nu} -  \frac{2 \xi_{\gamma}}{M} f^{\mu\lambda\nu}\pa_{\lambda}\,\hat{D}\big)A_{\nu}=0.
\eea

After a straightforward algebra we find that the free continuous spectrum associated with the equation of motion, Eq.  (\ref{EE02}), is governed by the following covariant dispersion relation:
\bea\label{EE03}
(k_{\gamma}^{2})^{2} - (2 \xi_{\gamma}/M)^{2} \big((n\cdot k_{\gamma})^{2} - n^{2} k_{\gamma}^{2}\big)^{3}=0
\eea
which was also derived in Ref. \cite{Reyes}.

\subsection{Modified propagations to electromagnetic waves}
The solutions to the above dispersion relation for the isotropic configuration, $i.e.$ when $n_{\mu} \equiv (1,\vec{0})$, is chosen to be purely time-like as investigated in Ref. \cite{Reyes}.{ This approach corresponds to a small subset of general LIV operators which preserves the rotational invariance. Such isotropic inertial frame must be specified because observer boosts to other frames destroy the rotational invariance. One natural choice for the preferred frame is the frame of the cosmic microwave background (CMB), but other choices are possible as is discussed in \cite{Kost_Mewes_01}.} From this isotropic configuration we generalize the dispersion relations associated with dimension$-n$ operators:
\bea\label{EE04}
E_{\gamma}^{2} - k_{\gamma}^{2} -  2 \lambda \xi^{(n)}_{\gamma} \frac{k_{\gamma}^{n}}{M^{n-2}} =0,\;\;\;\;\; k_{\gamma}\equiv |\vec{k}_{\gamma}|
\eea
with the two polarizations $\lambda =  \pm 1$. For $n=3$ we recover the cubic modifications reported in \cite{MP}. And, for $n=4,5,...$ we find new expressions due to the increase of the dimension of the LIV operator. 

Solving Eq. (\ref{EE04}) for the energy, we obtain the frequency solutions
\bea\label{EE05}
E_{\gamma}= k_{\gamma}\sqrt{1 + 2 \lambda\xi^{(n)}_{\gamma} \big(k_{\gamma}/M \big)^{n-2}}.
\eea

The dispersion relation (\ref{EE05}) leads to a modified speed of light for a photon with momentum $k_{\gamma}$:
\bea\label{EE06}
v^{(\gamma)}_{g} \equiv \frac{\pa E_{\gamma}}{\pa k_{\gamma}} = \frac{1 + n \lambda \xi^{(n)}_{\gamma}\big(k_{\gamma}/M \big)^{n-2} }{\sqrt{1 + 2 \lambda \xi^{(n)}_{\gamma}\big(k_{\gamma}/M \big)^{n-2} }}.
\eea
This dispersion relation leads to rotations of the polarization of linearly polarized photons during their propagation (see, $e.g.$, Refs. \cite{Matt} and  \cite{Passos:2016bbc}). Notice that for $n \ge 3$ the speed $v_{\gamma^{(\lambda= -)}}$ can exceed the speed of light introducing problems of causality (see also Ref. \cite{Reyes}).

On the other hand, the phase velocity can be obtained with $v^{(\gamma)}_p=\frac{E_\gamma}{k_\gamma}$:
\bea\label{eqe20}
v^{(\gamma)}_{p} = \sqrt{1 + 2 \lambda\xi^{(n)}_{\gamma} \big(k_{\gamma}/M \big)^{n-2}}.
\eea
Notice that the phase and group velocities are related through Rayleigh's formula: $v_{p}/v_{g}= 1 - \big(E/v_{p}\big)\big(d v_{p}/d k \big)$. Thus, from Eq.~(\ref{EE06}) and Eq.~(\ref{eqe20}) we find
\bea\label{eqe20.1}
\frac{v^{(\gamma)}_{p} - v^{(\gamma)}_{g}}{v^{(\gamma)}_{g}}&=& -  \frac{(n-2) \lambda \xi^{(n)}_{\gamma}\big(k_{\gamma}/M \big)^{n-2}}{\sqrt{1 + n \lambda \xi^{(n)}_{\gamma}\big(k_{\gamma}/M \big)^{n-2}}}\nonumber\\&\approx &
-(n - 2) \lambda \xi^{(n)}_{\gamma}\bigg(\frac{k_{\gamma}}{M} \bigg)^{n-2}.
\eea
In the subluminal case, $\lambda = -1$ we have that $v_{p} > v_{\g}$, a normal dispersion
medium. And, in the superluminal case, $\lambda=+1$ this implies at $v_{\gamma}> v_{p}$, an anomalous medium (from an influence of anisotropic effects). Therefore, we can conclude that a model truly isotropic must be attributed only to 
subluminal case. This is important to phenomenological analyses.

\section{The extended linearized gravity: gravitational sector }\label{sec03}

\subsection{The Model}
In analogy to the above electromagnetic case,  we consider the following LIV extension to Fierz-Pauli action proposed in Ref. \cite{GravPassos}:
\bea\label{pi01}
S_{g} &=&-\frac{(M_{1})^{3} \xi_{g}}{2}\int d^{4}x\, f^{\mu\lambda\nu} h_{\rho\mu}\pa_{\lambda}h^{\rho}_{\,\nu},\;\; f^{\mu\lambda\nu} \equiv \varepsilon^{\alpha\mu\lambda\nu}n_{\alpha}
\eea
where $\xi_{g}$ is a dimensionless parameter, $M_{1}$ is the mass scale where LIV in the gravitational sector become pronounced. Here the $h_{\mu\nu}$ is a second rank symmetric tensor characterizing weak metric fluctuations ($h_{\mu\nu} = g_{\mu\nu} - \eta_{\mu\nu}$, where $g_{\mu\nu}$ is the metric tensor of the curved space, $\eta_{\mu\nu} = {\rm diag}(-1, +1, +1, +1)$ is the metric tensor of the flat space and $ h = \eta^{\mu\nu}h_{\mu\nu}$ is the trace of $h_{\mu\nu}$ \cite{Hinterbichler:2011tt}).

Notice that under gauge transformations $ \delta h_{\mu\nu} = \pa_{\mu} \xi_{\nu} + \pa_{\nu} \xi_{\mu}$ for a spacetime dependent gauge parameter $\xi_{\mu}(x)$ the action (\ref{pi01}) implies in the following variation: $\delta {\cal L}_{g}\sim f^{\mu\lambda\nu}\xi_{\mu}\pa_{\lambda}\pa_{\rho}h^{\rho}_{\,\nu}$ which does not vanish in general, so that the action $S_{g}$ is not gauge invariant. This issue can be investigated using the Stuckelberg formalism to a massive spin-two field  \cite{Hinterbichler:2011tt}.  {The presence of explicit LIV terms on the gravity sector leads to an apparent inconsistency as the diffeomorphism is broken and $\nabla_{\mu}T^{\mu\nu}\neq 0$. This inconsistency can be solved if the  breaking of the Lorentz invariance occurs spontaneously. These results were shown in Ref. \cite{Kostelecky:2003fs}.  However, in Ref. \cite{Bluhm:2014oua}, it was shown that the former result of Ref. \cite{Kostelecky:2003fs} was too strong and that the presence of explicit LIV terms are also permitted as long as some conditions are satisfied. It was also shown that massive gravity satisfies such conditions.}

Following last section, we rewrite the Eq.~(\ref{pi01}) in terms of a power series such as in the electromagnetic case. Again we replace the LIV operator to the following power series

\bea\label{pi02}
\sum_{l=1,3,...}\frac{\xi_{g_{l}}}{(M_{1})^{l - 2}} (\Pi^{\mu\nu})^{l}&=&
\xi_{g_{1}} M_{1} \Pi^{\mu\nu} + \frac{\xi_{g_{3}}}{M_{1}}\Pi^{\mu\nu}\hat{D}+ ...
\eea
such that
\bea\label{pi03}
S_{g}\to \hat{S}_{g} =-\frac{(M_{1})^{2}}{2}\int d^{4}x\,\Big[ M_{1} \xi_{g_{1}} f^{\mu\lambda\nu} h_{\rho\mu}\pa_{\lambda}h^{\rho}_{\,\nu} + 
\frac{\xi_{g_{3}}}{M_{1}} f^{\mu\lambda\nu} h_{\rho\mu}\pa_{\lambda} \hat{D}\, h^{\rho}_{\,\nu}+ ...\Big].
\eea

Therefore, we also derive higher-dimensional operators as extra terms of power series expansion which lead to cubic modifications of the dispersion relations of gravitational waves.  Although the restriction on the gauge invariance, the extra contribution in Eq.(\ref{pi03})  satisfies all the Myers-Pospelov criteria to construct LIV higher derivative operators. 

In this case we consider the second contribution in Eq.(\ref{pi03}) as a modification in the dynamics of the massive Fierz-Pauli action. Here we also analyze the dynamics associated with the dispersion relation of the gravitational waves.  Considering the following Lagrangian (in our case, for brevity, $ \xi_{g_{3}} \equiv\xi_{g}$):
\bea\label{eq01}
{\cal L}_{g}&=& \frac{(M_{1})^{2}}{2}\Big[\frac{1}{2}\pa_{\lambda}h_{\mu\nu}\pa^{\lambda}h^{\mu\nu} + \pa_{\mu}h_{\nu\lambda}\pa^{\nu}h^{\mu\lambda} -\pa_{\mu}h^{\mu\nu}\pa_{\nu}h+\frac{1}{2}\pa_{\lambda}h\pa^{\lambda}h +\nonumber\\&&
 \frac{1}{2}m^{2}_{g}\big(h_{\mu\nu}h^{\mu\nu} - h^{2} \big) - \frac{\xi_{g}}{M_{1}} f^{\mu\lambda\nu} h_{\rho\mu}\pa_{\lambda}\hat{D} \,h^{\rho}_{\,\nu}\Big].
\eea
We will take this Lagrangian since it can describe a massive spin-two LIV theory. Notice that there is no gauge symmetry due to a mass term and the $-1$ coefficient in $\big(h_{\mu\nu}h^{\mu\nu} - h^{2} \big)$ is dubbed as {\it Fierz-Pauli tuning}. In this paper, we are interested in the phenomenological aspects associated with Eq. (\ref{eq01}).

{Notice that our model, Eq.  (\ref{eq01}), has two potentially harmful terms, namely the presence of the graviton mass, $m_{g}$, and a fixed background field, $n^{\mu}$, which can give rise to inconsistencies between geometry and dynamics \cite{Kostelecky:2003fs,Bluhm:2014oua}. We now discuss how these inconsistencies can be evaded. For the massive gravity terms without the background field, we can proceed twofold: first, it was show in Ref.  \cite{Bluhm:2014oua} that the massive gravity automatically avoids these inconsistencies. Second, one can think that the massive field is not fundamental, $i.e.$ it only exists after some condensation process occurs. It can be shown that this mass term induced by a condensate can appear without offending the original gauge symmetry \cite{JT}, in conformity  with the Elitzur theorem \cite{Elitzur:1975im}  which states that there is no spontaneously symmetry breaking. To  dwell upon the background field terms, we again can proceed twice. First, these background field can appear after a spontaneous break of Lorentz symmetry, and this evades the negative result of Ref. \cite{Kostelecky:2003fs}. Second, we have omitted the the kinetic terms for the background field, these terms can trigger the spontaneous violation of Lorentz symmetry. This full Lagrangian has no inconsistency as $\nabla_{\mu}T^{\mu\nu}=0$, and for an explicit worked example, see Ref. \cite{Rougemont}. }

{
The equations of motion from (\ref{eq01}) are given as
\bea\label{eq02A}
G^{\mu\nu} + C^{\mu\nu} +  M^{\mu\nu}=0
\eea
where
\begin{subequations}
	\bea\label{eq02B}
	G^{\mu\nu}=\Box h^{\mu\nu} - \pa_{\lambda}\pa^{\mu}h^{\lambda\nu} - \pa_{\lambda}\pa^{\nu} h^{\lambda\mu} + \eta^{\mu\nu} \pa_{\lambda}\pa_{\sigma}h^{\lambda\sigma} +\pa^{\mu}\pa^{\nu}h- \eta^{\mu\nu}\Box h,
	\eea
	\bea\label{eq02B1}
	C^{\mu\nu} =  - \big(\xi_{g}/M_{1}\big)f^{\mu\lambda\alpha}\pa_{\lambda} 
	\hat{D}\,\eta^{\beta\nu}h_{\alpha\beta}-\big(\xi_{g}/M_{1}\big)f^{\nu\lambda\beta}\pa_{\lambda} \hat{D}\,\eta^{\alpha\mu}h_{\alpha\beta},
	\eea
	\bea\label{eq02B2}
	M^{\mu\nu}=  m_{g}^{2}\big(h^{\mu\nu} - \eta^{\mu\nu}h\big).
	\eea
\end{subequations}
All the terms above are symmetric in the free indices. The consistency condition requiring the vanishing divergence of the above field equations can
be directly verified. The linearized Einstein tensor is naturally divergenceless,
i.e. $\pa_{\mu}G^{\mu\nu}=0$. The divergence of the LIV term becomes
\bea\label{eq02B3}
 \pa_{\mu} C^{\mu\nu}= - \big(\xi_{g}/M_{1}\big)n_{\mu\sigma}\varepsilon^{\sigma\nu\lambda\beta}\pa_{\lambda} \hat{D}\,\eta^{\alpha\mu}h_{\alpha\beta},
\eea
where $n_{\mu\sigma}= \pa_{\mu}n_{\sigma}$. If we consider that our fixed background field is given by $n_{\sigma}=\pa_{\sigma}\theta$, with $\theta$ being a non-dynamical scalar field, we find that 
$\pa_{\mu} C^{\mu\nu}=0$, by symmetry considerations. So we have only the following constraint:
\bea\label{c01}
\pa_{\mu}h^{\mu\nu}=\pa^{\nu}h
\eea
which we insert this back into the above equations of motion to get
 \bea\label{eq03}
 \Box h^{\mu\nu} - \pa^{\mu}\pa^{\nu}h + m_{g}^{2}(h^{\mu\nu} - \eta^{\mu\nu}h)-  \frac{\xi_{g}}{M_{1}}f^{\mu\lambda\alpha}\pa_{\lambda} 
 \hat{D}\,\eta^{\beta\nu}h_{\alpha\beta}-\frac{\xi_{g}}{M_{1}}f^{\nu\lambda\beta}\pa_{\lambda} \hat{D}\,\eta^{\alpha\mu}h_{\alpha\beta} =0.
 \eea
Taking the trace of above equation this implies $h=0$ and as a consequence of (\ref{c01}) one finds $\pa_{\mu}h^{\mu\nu}=0$. The Eq.(\ref{eq03}) reduces to the following form:
\bea\label{eq041}
\Big[\big(\Box + m_{g}^{2}\big) \eta^{\alpha\mu}\eta^{\beta\nu} - \frac{\xi_{g}}{M_{1}}f^{\mu\lambda\alpha}\pa_{\lambda} 
\hat{D}\,\eta^{\beta\nu}-\frac{\xi_{g}}{M_{1}}f^{\nu\lambda\beta}\pa_{\lambda} \hat{D}\,\eta^{\alpha\mu}\Big]h_{\alpha\beta} =0.
\eea
Notice that the divergenceless of $G^{\mu\nu}$ and the above conditions on $h^{\mu\nu}$ reduce the number of degrees of freedom in (\ref{eq02A}) to only 2 degrees of freedom governed by equation (\ref{eq041}) as expected. 

In order to find the dispersion relation we get Eq.(\ref{eq041}) in the momentum
space by considering the Ansatz: $h_{\alpha\beta}= \epsilon_{\alpha\beta}\, {\rm exp}(i k_{\mu}x^{\mu})$. Thus, we find
\bea\label{eq041A}
\Big[\big(k_{g}^{2} - m_{g}^{2}\big) \eta^{\alpha\mu}\eta^{\beta\nu} - \frac{i\xi_{g}}{M_{1}}f^{\mu\lambda\alpha}k_{\lambda} 
\hat{D}(k)\,\eta^{\beta\nu}-\frac{i\xi_{g}}{M_{1}}f^{\nu\lambda\beta}k_{\lambda} \hat{D}(k)\,\eta^{\alpha\mu}\Big]\epsilon_{\alpha\beta} =0
\eea
where $\hat{D}(k) = \big((n\cdot k_{g})^{2} - n^{2}k_{g}^{2}\big)$. Now, we apply $\big(\big(k^{2} - m_{g}^{2}\big) \eta_{\sigma\mu}\eta_{\tau\nu} - \frac{i\xi_{g}}{M_{1}}f_{\mu\rho\sigma}k^{\rho} 
\hat{D}(k)\,\eta_{\tau\nu}-\frac{i\xi_{g}}{M_{1}}f_{\nu\rho\tau}k^{\rho} \hat{D}(k)\,\eta_{\sigma\mu}\big)$ in the above equation to obtain
\bea\label{eq05A}
\Big[\big(k_{g}^{2} - m^{2}_{g}\big)^{2} - \big(2\xi_{g}/M_{1}\big)^{2}\big((n\cdot k_{g} )^{2} - n^{2}k_{g}^{2}\big)^{3}\Big] \delta^{\alpha}_{\sigma}\delta^{\beta}_{\tau}\,\epsilon_{\alpha\beta}=0
\eea
where we have used the conditions $k^{\alpha}\epsilon_{\alpha\beta}=n^{\alpha}\epsilon_{\alpha\beta}=0$, $\epsilon_{\alpha\alpha}=0$. The free continuous spectrum associated with Eq.~(\ref{eq05A}) is given by the following dispersion relation
\bea\label{eq05}
\big(k_{g}^{2} - m^{2}_{g}\big)^{2} - \big(2\xi_{g}/M_{1}\big)^{2}\big((n\cdot k_{g} )^{2} - n^{2}k_{g}^{2}\big)^{3}=0.
\eea
 Notice that for $m_{g}=0$, the Eq.~(\ref{eq05}) is equivalent to the Eq.~(\ref{EE03}), i.e., the electromagnetic case, given $e.g.$ in Ref. \cite{Reyes}.
}

\subsection{Modified propagations to gravitational waves}
Moreover, we study the solutions for the dispersion relation given by Eq. (\ref{eq05}) in the isotropic configuration, that is, for  $n_{\mu}=(1,\vec{0})$ chosen to be purely time-like, for dimension$-n$ operators. 
Thus, we have 
 \bea\label{Eq06}
E_{g}^{2} - k_{g}^{2} - m_{g}^{2} -  2 \lambda \xi^{(n)}_{g} \frac{k_{g}^{n}}{M_{1}^{n-2}} =0,\;\;\;\;\; k_{g}\equiv |\vec{k}_{g}|
\eea
 with the two polarizations $\lambda =  \pm 1$.  
 
 Solving the Eq.~(\ref{Eq06}) for $E_{g}$ we find the frequency solutions
 \bea\label{eq07}
 E_{g}= \sqrt{ k^{2}_{g} \big(1 + 2 \lambda\xi^{(n)}_{g} \big(k_{g}/M_{1} \big)^{n-2}\big) + m^{2}_{g}}.
 \eea
Notice also that the solutions correctly reproduce the usual ones in the limit $\xi_{g}\to 0$ given in Ref. \cite{Hinterbichler:2011tt}.
We assume here the graviton velocity, $v_{g}$, is given by the group velocity determined from the dispersion relation
(\ref{eq07}), that is
 \bea\label{eq08}
 v^{(g)}_{g}\equiv\frac{\pa E_{g}}{\pa k_{g}}=  \frac{1 +n \lambda \xi^{(n)}_{g} \big(k_{g}/M_{1} \big)^{n-2}}{\sqrt{ 1 +2 \lambda\xi^{(n)}_{g} \big(k_{g}/M_{1} \big)^{n-2} + (m_{g}/k_{g})^{2}}}. 
 \eea
On the other hand, the phase velocity can be obtained with $v_{p}=\frac{E_{g}}{k_{g}}$: 
 \bea\label{eqg25.2}
 v^{(g)}_{p}= \sqrt{ 1 + 2 \lambda\xi^{(n)}_{g} \big(k_{g}/M_{1} \big)^{n-2} + \big(m_{g}/k_{g}\big)^{2}}.
 \eea
 The relation between the Eq.(\ref{eq08}) and Eq.(\ref{eqg25.2}) from Rayleigh's formula leads us to
 \bea\label{eqg26}
 \frac{v^{(g)}_{p} - v^{(g)}_{g}}{v^{(g)}_{g}} &=&   \frac{- \lambda(n - 2) \xi^{(n)}_{g} \big(k_{g}/M_{1}\big)^{n-2} + \big(m_{g}/k_{g}\big)^{2} }{ 1 + n \lambda\xi^{(n)}_{g} \big(k_{g}/M_{1} \big)^{n-2}}\nonumber\\&\approx &
\bigg( \frac{m_{g}}{k_{g}}\bigg)^{2} - \lambda (n - 2) \xi^{(n)}_{g} \bigg(\frac{k_{g}}{M_{1}}\bigg)^{n-2} - n \lambda \xi^{(n)}_{g} \bigg(\frac{k_{g}}{M_{1}} \bigg)^{n-4}\bigg(\frac{m_{g}}{M_{1}}\bigg)^{2}.
\eea
For $\lambda=-1$ this imply that the $v_{p} > v_{g}$, a normal dispersion medium. And for $\lambda=+1$ we analyse two cases: (${\it i}$)  $\xi^{(n)}_{g} \big(k_{g}/M_{1} \big)^{n-2}\big((n-2) + n \big(m_{g}/k_{g}\big)^{2} \big) > \big(m_{g}/k_{g}\big)^{2}$ this imply that $v_{g} > v_{p}$, an anomalous medium and for (${\it ii}$) $\xi^{(n)}_{g} \big(k_{g}/M_{1} \big)^{n-2}\big((n-2) + n \big(m_{g}/k_{g}\big)^{2} \big) < \big(m_{g}/k_{g}\big)^{2}$ this imply that $v_{g} < v_{p}$, an normal dispersion medium. 

 
\section{Phenomenological Aspects}\label{sec04}
 In the following we consider some expressions on photons and massive gravitons velocities to impose the upper bounds for the $\xi_{g},\;\xi_{\gamma} -$ LIV. To do this, we use the Fermi Gamma-Ray Burst Monitor (GMB)-LIGO  observations associated with a transient source based in the following measured of time arrival delay: $\Delta t\sim 0.40\,{\rm s}$ between the gamma-ray burst and the gravitational wave \cite{Fermi01}. {Recently, a LIV gravity sector was introduced to investigate its effects in gravitational waves and the behavior of gravity in short-range scales \cite{Bailey}, however they do not consider the simultaneous LIV of the electromagnetic and gravity sectors.}

 
\subsection{Graviton propagation}
Let us now derive a relation to parameters which controls the LIV  in the massive gravitons sector, $\xi_{g}$ from associate group velocity. For $\lambda = -1$, we find the folling expression: 
\bea\label{eq09}
v^{(g)}_{g} \approx 1 - \frac{1}{2}\bigg( \frac{m_{g}}{k_{g}}\bigg)^{2} -  (n - 1)  \xi^{(n)}_{g} \bigg(\frac{k_{g}}{M_{1}} \bigg)^{n-2}  +\;  \frac{1}{2}  n   \xi^{(n)}_{g} \bigg(\frac{k_{g}}{M_{1}} \bigg)^{n-4}\bigg(\frac{m_{g}}{M_{1}}\bigg)^{2}.
\eea
 In the limit $M_{1} \gg m_{g}$, the Eq.~(\ref{eq09}) takes the form
  \bea\label{eq09a}
 v^{(g)}_{g} \approx 1 - \frac{1}{2}\bigg( \frac{m_{g}}{k_{g}}\bigg)^{2} -  (n - 1)  \xi^{(n)}_{g} \bigg(\frac{k_{g}}{M_{1}} \bigg)^{n-2}.
 \eea
From Eq.~(\ref{eq09a}) we can obtain two possibles regimes to difference between the speed
of gravitons, $v^{(g)}_{g}$  and the speed of light, $c$. In the absence of LIV, first we have
\bea\label{eq10a}
\Delta v_{g} = c - v^{(g)}_{g}\Big|_{\xi_{g}=0}  \approx \frac{1}{2}\bigg( \frac{m_{g}}{k_{g}}\bigg)^{2}.
\eea
Considering that the LIGO Scientific and Virgo Collaborations \cite{GW01}  pointed out that the signal of gravitational wave event GW150914 is peaked at $ \nu=150 \,{\rm Hz}$, then the estimated energy of gravitons is $k_{g} = h\nu \approx  6.024 \times 10^{-13}\,{\rm eV}$ (with $h= 4.136 \times 10^{-15} {\rm eV\cdot s}$.) Moreover, they also found an upper bound for the mass of gravitons $m_{g} \lesssim 1.20 \times 10^{-22} {\,\rm eV}$.  Therefore, Eq.~(\ref{eq10a}) turns out to be
\bea\label{eq11a}
\Delta v_{g} \lesssim 2.0 \times 10^{-20}.
\eea
It is clear that the difference between the speed of gravitons and the speed of light is very small.
Second, to massless gravitons, difference between the modified speed of gravitons and the
speed of light is
\bea\label{eq12a}
\Delta v^{\prime}_{g} = c - v^{(g)}_{g}\Big|_{m_{g}=0} \approx (n - 1)  \xi^{(n)}_{g} \bigg(\frac{k_{g}}{M_{1}} \bigg)^{n-2}
\eea
Note that we can set improved upper limits on the LIV  parameter,  by setting (\ref{eq11a})  as an upper bound to (\ref{eq12a}), such that
\bea\label{eq13a}
\xi^{(n)}_{g} \lesssim \frac{(0.73) \times 10^{(13n - 44)} }{(6.024)^{n}(n - 1)} {\Big(\frac{M_{1}}{\rm eV}\Big)^{n-2}}.
\eea
For the case $n=3$, we have
\bea\label{eq14a}
\xi_{g} \lesssim \big(1.66\, M_{1}\big)\times 10^{-8} \big(\rm eV\big)^{-1}
\eea
which corresponds to a mass-scale dependent parameter. Particularly, if we use 
$M_{1} \sim 10^{5}{\,\rm eV}$ \cite{Ellins}, we obtain that this upper bound is $\xi_{g} \sim 10^{-3}$, which can be relevant phenomenologically.{
In terms of length scale, 
Eq.(\ref{eq14a}) can be rewritten as 
\bea\label{eq14aa}
{\xi}_{g} =M_1\bar{\xi}_g,\qquad \bar{\xi}_{g} \lesssim 3.27\,\times 10^{-15}{\rm\, m}
\eea
where $ \bar{\xi}_{g} $ is compatible with the bounds obtained in \cite{Kos_Grav_01} to dimension-five operator in modified gravitation. 

}

 \subsection{Time delay between the flight of photons and gravitons}

The difference $\Delta t = t_{g} - t_{\gamma}$ between the propagation of the gravitational and electromagnetic waves is given by  \cite{Vincenzo}
\bea\label{TC5}
\Delta t = \Delta t_{a} - (1 + z) \Delta t_{e}
\eea
where $\Delta t_{a}$ (measured quantity) is the arrival delay observed at the Earth  and $\Delta t_{e}$ (unknown quantity) is the emission delay at the source with redshift $z$. Here we assume that  $\Delta t_{e}=0$ (the simultaneous emission of gravitational and electromagnetic waves) to derive constraints on LIV by velocities of the gravitons and photons. 

In general, the group velocities given from Eq.~(\ref{EE06}) (for photons) and from Eq.~(\ref{eq08}) (for gravitons) implies that photons and gravitons of
different wave vectors $k_{1}$ and $k_{2}$ travel at slightly different speeds. Let us first assume that there are no anisotropic effects associated with photons, so that $\lambda=-1$. And after that there are no LIV effects associated with gravitons, so that $\xi_{g}=0$. Then, upon propagation on a cosmological
distance $d$, the effect of the energy dependence of the photons and gravitons group velocities produces a time delay (for spatially flat Universe, $\Omega_{k}=0$):
\bea\label{TC7}
\Delta t_{a}=   H_{0}^{-1}\bigg((n - 1)\; \xi_{\gamma}^{(n)}  \bigg(\frac{k_{\gamma}}{M}\bigg)^{n-2} - \frac{1}{2}\frac{m_{g}^{2}}{k_{g}^{2}} \bigg) \int_{0}^{z} \frac{d z^{\prime}}{\sqrt{\Omega_{m} (1 + z^{\prime})^{3} + \Omega_{\Lambda} }}.
\eea
where $H_{0}= 67.8 {\rm\, Km\,  (s\,Mpc})^{-1}$ is the Hubble constant ($H_{0}^{-1}=4.55 \times 10^{17} {\rm\, s}$) with
$\Omega_{m}$ and $ \Omega_{\Lambda}$ being the matter and dark energy density parameters, respectively. As warned in \cite{liberati01}, the Eq.~(\ref{TC7}) has been constructed such that all effects associated with birefringent theories are excluded. 

From  Eq.~(\ref{TC7}) we get
\bea\label{TC8}
\xi^{(n)}_{\gamma} = \frac{1}{(n - 1)}\bigg(\frac{M}{k_{\gamma}}\bigg)^{n-2} \Bigg[\frac{m_{g}^{2}}{2 k_{g}^{2}} + \frac{ \Delta t_{a} }{ H_{0}^{-1}} \bigg(\int_{0}^{z} \frac{d z^{\prime}}{\sqrt{\Omega_{m} (1 + z^{\prime})^{3} + \Omega_{\Lambda} }} \bigg)^{-1}\Bigg]
\eea
Now performing the integral for $\Omega_{m}=0.31$, $\Omega_{\Lambda}=0.69$ at a redshift $z=0.09$ and using the previously assumed values for $\Delta t_{a} =0{.}40$ s, $k_{g} \sim 6.024 \times 10^{-13}$ eV and  $k_{\gamma}\sim 5.00 \times 10^{4} {\rm eV}$ being the photon energies for transient source measured by GMB \cite{Fermi01}, we obtain
\bea\label{TC801}
\xi^{(n)}_{\gamma} \lesssim \frac{\big(1.10\big)\times 10^{-(4n+9)}}{(n-1) (5.0)^{n-2}} \bigg(\frac{M}{\rm{eV}}\bigg)^{n-2}
\eea
Notice that for $n=3$,  we have
\bea\label{TC09}
\xi^{(n=3)}_{\gamma} \lesssim (1.10\,M) \times 10^{-22} (\rm{eV})^{-1}
\eea
which also corresponds to a mass-scale dependent parameter. In particular, inserting  $M\sim10^{28}$ eV, as suggested, in Refs. \cite{Albert:2007qk,Ellins} into Eq. (\ref{TC09}), $\xi^{(n=3)}_{\gamma}\sim10^{6}$ is gotten. Hence for this very large $M$ value, our result is not suitable for any realistic phenomenology. One then concludes that either the value of $M$ has to be modified (see also Ref. \cite{Passos:2016bbc}, where another energy scale, namely the Ho\v{r}ava-Lifshitz one is also introduced to give more realistic bounds) or this term cannot be present in the description of a LIV effective theory. {We can also compare the result given by Eq.~(\ref{TC09}) with other astrophysics bounds obtained by different models. First we rewrite Eq.~(\ref{TC09}) as
\bea\label{TC09a}
{\xi}_{\gamma}= \bar{\xi}^{(n=3)}_{\gamma}M, \qquad \bar{\xi}_{\gamma}\lesssim& 1.10\, \times 10^{-13} (\rm{GeV})^{-1}
\eea	 
Now, we consider the astrophysical birefringence test derived from dimension-5 operators whose bounds obtained are represented from collected data in \cite{Kost_Russel_01}. Using the Table $\rm D15$ associated with nonminimal electromagnetic photon sector, we find values that run from $|k^{(5)}_{(V)00}|<1\times 10^{-34}(\rm{GeV})^{-1}$ to $|k^{(5)}_{(V)00}|<5.1\times 10^{-24}(\rm{GeV})^{-1}$ and several bounds around $|k^{(5)}_{(V)00}|\sim 1\times 10^{-20}(\rm{GeV})^{-1}$ can also be found for CMB polarization. For the sake of comparison we have 
\bea\label{TC09b}
\frac{|k^{(5)}_{(V)00}|}{\bar{\xi}_{\gamma}}\lesssim \; 10^{-11}
\eea	
for the larger aforementioned astrophysical birefringence bound, and 
\bea\label{TC09bbb}
\frac{|k^{(5)}_{(V)00}|}{\bar{\xi}_{\gamma}}\lesssim \; 10^{-7}
\eea	
for the CMB polarization bound. On the other hand, the ration between the smallest astrophysical birefringence bound above $|k^{(5)}_{(V)00}|$ and the CMB polarization bound is $\lesssim 10^{-14}$.  This means that the range of deviations of our bounds seems to be in agreement with the deviations among others well-known bounds in the literature.
}

To complete our phenomenological analysis let us compare Eq.~(\ref{eq13a}) with Eq.~(\ref{TC801}). As a consequence, we find the following relationship
\bea\label{TC10}
\frac{\xi^{(n)}_{g}}{\xi^{(n)}_{\gamma}}= \big( 8.30\big)^{n} \big(2.65\big)\times 10^{17n - 35}\bigg(\frac{M_{1}}{M}\bigg)^{n-2}.
\eea
 Therefore, for $(n=3)$, we find,
 \bea\label{TC11}
\frac{\xi^{(n=3)}_{g}}{\xi^{(n=3)}_{\gamma}}= 1.50 \times 10^{14} \bigg(\frac{M_{1}}{M}\bigg).
\eea
Notice that if $M\gg M_{1}$, the above quantity may lead to a realistic constraint. 
Now we calculate a relation between the light and graviton velocities.  From Eq.~(\ref{EE06})  we obtain the following expression to effective velocity of subluminal photons:
\bea\label{TC12}
\Delta v_{\gamma}= c - v^{(\gamma)}_{g} = \big(n - 1 \big) \xi^{(n)}_{\gamma} \bigg(\frac{k_{\gamma}}{M} \bigg)^{n-2}.
\eea
At $(n=3)$,  we can combine Eqs.~(\ref{TC09})--(\ref{TC12}) to find
\bea\label{TC12.2}
\Delta v_{\g}\lesssim 1.10\,\times 10^{-17} 
\eea
{
Notice that from the Eqs.~(\ref{eq11a}) and (\ref{TC12.2}) we find
\bea\label{TC12.3}
\frac{\Delta v_{g}}{\Delta v_{\gamma}}\lesssim 1.82\times 10^{-3} .
\eea
The above result can be seen as tiny difference between the propagation of fields of spin 1 and spin 2 in a possible quantum spacetime setup. This can be expected, as there are different  cut-off energy scales where the Lorentz violating effects become prominent for these two kinds of fields. Furthermore, we can also obtain
\bea\label{TC12.4}
\Delta v_{\gamma} - \Delta v_{g} \lesssim 10^{-17}.
\eea
which is in accord with the bounds found in \cite{Ellins} --- see also  \cite{Gia}.
}

\section{Conclusions}\label{sec05}
In this work, we  analyze the LIV effects from electromagnetic and gravitational higher derivative operators using the Myers-Pospelov approach to obtain LIV effective theories. First, we extend the electromagnetic and massive gravitational actions to include LIV higher-order derivative terms. Then we compute the equations of motion, the dispersion relations for these sectors and the photon and graviton velocities. Assuming that the same process that generated the detected gravitational waves also emits electromagnetic waves, also detected by other means,  bounds for the LIV parameters for electromagnetic, $\xi_{\gamma}$, and massive gravitational, $\xi_{g}$, sectors are obtained for three approaches, namely,  luminal photons, time delay of flight and the difference between photon and graviton velocities. For the first two approaches, there is a dependence of $\xi_{g}$ and $\xi_{\gamma}$ on the respective mass scales $M_{1}$ and $M$, where the LIV effects become relevant. Using the value for $M_{1}$ obtained in Ref. \cite{Ellins}, it is gotten that $\xi_{g}\sim10^{-3}$, and this is expected to be phenomenological relevant. For the time delay of flight approach and the value of $M$ given in Refs. \cite{Albert:2007qk,Ellins}, it is found that $\xi_{g}\sim10^{6}$, which cannot represent any realistic LIV scenario. However, the ratio between $\xi_{g}$ and $\xi_{\gamma}$, Eq. (\ref{TC11}), can be made phenomenological relevant if $M \gg M_{1}$, which is satisfied, even if we consider that $M$ has to be changed to make $\xi_{\gamma}\sim 1$ in Eq. (\ref{TC09}). Finally, the ratio between the difference of photon and graviton velocities in relation to the light speed were computed showing the differences between the behavior of spin 1 and 2 fields in a possible quantum spacetime setup.


{\acknowledgments} We would like to thank to CNPq for partial financial support. C. A. D. Z. is thankful for the kindness and hospitality of the Physics Department at Federal University of Campina Grande, where part of this work was carried out. {The authors would like to thank Quentin Bailey for some useful comments on Sec. \ref{sec02} and Sec. \ref{sec03}.}

\eject

\end{document}